\documentclass[12pt]{article}
\usepackage{amsmath}
\usepackage{cite}
\usepackage[dvipdfmx]{graphicx}
\begin{document}    
\begin{flushright}
KANAZAWA-18-05\\
September, 2018
\end{flushright}
\vspace*{1cm}

\renewcommand\thefootnote{\fnsymbol{footnote}}
\begin{center} 
  {\Large\bf An extension of the SM based on effective Peccei-Quinn
    Symmetry}
\vspace*{1cm}

{\Large Daijiro Suematsu}\footnote[1]{e-mail:
~suematsu@hep.s.kanazawa-u.ac.jp}
\vspace*{0.5cm}\\

{\it Institute for Theoretical Physics, Kanazawa University, 
Kanazawa 920-1192, Japan}
\end{center}
\vspace*{1.5cm} 

\noindent
{\Large\bf Abstract}\\
Peccei-Quinn (PQ) mechanism based on a chiral global $U(1)$ symmetry 
is considered to be a simple and elegant solution for strong $CP$ problem. 
Fact that the mechanism could be experimentally examined 
through the axion search makes it much more interesting and recently
it causes a lot of attention again.
However, it is also known that the mechanism is annoyed by two serious 
problems, that is, a domain wall problem and goodness of global 
symmetry. Any global symmetry is considered not to be exact due to 
the quantum effect of gravity.  
In this paper, we consider a solution to these problems, in which 
quark mass hierarchy and mixing, neutrino mass generation 
and existence of dark matter are closely related.
In our solution,  PQ symmetry is assumed to be induced 
through symmetry breaking at an intermediate scale
of a local  U(1) symmetry, and a global U(1) symmetry which plays a role of
Froggatt-Nielsen symmetry . 
In the lepton sector, a remnant of the PQ symmetry controls neutrino mass 
generation and dark matter existence.
 
\newpage
\setcounter{footnote}{0}
\renewcommand\thefootnote{\alph{footnote}}

\section{Introduction}
Strong $CP$ problem is one of serious problems in the standard model (SM),
which is suggested by an experimental bound of the electric dipole 
moment of a neutron \cite{strongcp}. The bound requires a fine tuning of 
 $O(10^{-10})$ for a parameter $\theta$ in QCD.
Invisible axion scenario based on a chiral global 
symmetry, which is called Peccei-Quinn (PQ) symmetry \cite{pq,ww}, 
is known to give a simple and elegant solution to it. 
Since it predicts the existence of a light and 
very weakly interacting pseudoscalar \cite{ksvz,dfsz}, 
this solution could be examined
experimentally. Moreover, it is known to present a good candidate for
cold dark matter (DM) under a suitable condition \cite{axiondm}.
Its experimental search is proceeded now.

On the other hand, the scenario has two fatal problems generally.
The first one is known as a domain wall problem \cite{dw}. 
Although PQ symmetry is explicitly broken to its subgroup $Z_N$
through the QCD instanton effect, the $Z_N$ is also spontaneously 
broken to its subgroup when PQ symmetry is spontaneously broken
by a vacuum expectation value (VEV) of scalar fields and quark condensates.
This brings about $N$ degenerate vacua, which are separated by topological 
defects called domain wall. Since the energy density of domain walls
dominates cosmological energy density of the Universe inevitably,
the Universe is over-closed contradicting to the observations.
It is known that the domain wall problem could be escaped for a 
non-degenerate vacuum which has $N=1$ \cite{soldw},
even if the cosmological inflation occurs 
before the PQ symmetry breaking.

The second one is related to goodness of the PQ symmetry.
The PQ symmetry is a global symmetry, which is used to be considered broken 
by the gravitational effect \cite{gbr}. If this breaking effect due to 
the gravity is larger than the QCD instanton effect, the PQ mechanism cannot 
solve the strong $CP$ problem. In order to escape this dangerous situation, 
such symmetry breaking operators caused by the gravity should be
forbidden up to dimension ten \cite{bs}.
There, the PQ symmetry is considered to be realized as an accidental 
symmetry induced by some gauge symmetry or a discrete symmetry, 
which satisfies such a constraint on its goodness. 
In such a direction, several works has been done by now \cite{accidsym}. 
 
In this paper, we propose a model which can escape these two problems
in invisible axion models \cite{ksvz,dfsz}.  
Although the SM has been confirmed by the discovery of the 
Higgs scalar \cite{higgs}, it cannot explain several experimental and 
observational data such as quark mass hierarchy and 
CKM mixing, neutrino masses and their large mixing \cite{pdg}, and 
also the existence of DM \cite{dm}. 
In the present model construction, we take account of these
problems also.\footnote{Model construction to explain these problems
  including the strong CP problem has been done in various articles
  \cite{bsmcp,fn-pq,s,s1}.}.
For this purpose, we impose $U(1)_g\times U(1)_{FN}$ on 
the model, where $U(1)_g$ is a gauge symmetry but $U(1)_{FN}$ is a global 
symmetry whose charge is flavor dependent. 
Then, the latter could play a role of
Froggatt-Nielsen symmetry \cite{fn}.
This symmetry is assumed to be spontaneously broken to 
PQ symmetry $U(1)_{PQ}$ at some intermediate scale.
We require that $U(1)_g$ guarantees
the goodness of $U(1)_{PQ}$ to be kept up to a consistent level
required by the strong $CP$ problem.
After the spontaneous breaking of $U(1)_{PQ}$, both a non-degenerate QCD 
vacuum and Yukawa couplings with desirable flavor structure are induced
in a quark sector \cite{s1}. 
In a leptonic sector, the scotogenic model \cite{ma} which connects 
the neutrino mass generation and the existence of DM is brought about 
as a low energy effective model.    

The remaining parts are organized as follows. In the next section,
we present a model by fixing the symmetry $U(1)_g\times U(1)_{FN}$ and
the field contents in the model. 
We discuss features of the model such as the symmetry 
breaking, the domain wall number, the goodness of the PQ symmetry and so on. 
In section 3, phenomenological features of this model are discussed, 
such as quark mass hierarchy and CKM mixing, neutrino mass generation, 
leptogenesis, DM abundance and so on. 
We summarize the paper in section 4.

\section{A model with $U(1)_g\times U(1)_{FN}$}
We start presenting a brief review of QCD vacuum degeneracy
in the PQ mechanism \cite{strongcp}.
If the $U(1)_{PQ}[SU(3)_c]^2$ anomaly takes a value $N$
for the PQ charge assignment of colored contents in the model,
the $U(1)_{PQ}$ transformation of the colored fermions shifts a 
 parameter $\bar\theta$ as \cite{pq} 
\begin{equation}
\bar\theta \rightarrow \bar\theta +2\pi N,
\end{equation}
where $\bar\theta$ is a coefficient of an effective term
$\frac{\bar\theta}{32\pi^2}F^a_{\mu\nu}\tilde F^{a\mu\nu}$ induced
by instantons and it is defined as $\bar\theta=\theta+
{\rm arg}({\rm det}~{\cal M})$. ${\cal M}$ stands for a quark mass matrix.
If the PQ symmetry is spontaneously broken by a VEV of 
a scalar field $S$, $\bar\theta$ behaves a dynamical variable corresponding
to a pseudo Nambu-Goldstone boson associated with this breaking,
which is called axion $a$ \cite{ww,ksvz,dfsz}. 
Since a period of $\bar\theta$ is $2\pi$ and potential for $\bar\theta$
can be represented by using a QCD scale $\Lambda_{\rm QCD}$ as 
\begin{equation}
V(\bar\theta)=\Lambda_{\rm QCD}^4(1-\cos\bar\theta), 
\label{axionpot}
\end{equation}
this potential for $\bar\theta$ has $|N|$-fold degenerate minima.
The axion $a$ is fixed as
$a\equiv |\langle S\rangle|\frac{\bar\theta}{|N|}$ which 
is defined at a region $[0,2\pi)$.
This requires that axion decay constant $f_a$ 
should be identified as $f_a|N|=|\langle S\rangle|$.

\begin{figure}[t]
\begin{center} 
\begin{tabular}{c||ccccccccc}\hline
 &$Q_{L}^{(1)}$ &$Q_{L}^{(2)}$ & $Q_{L}^{(3)}$& $Q_{R}^{(1)}$ &$Q_{R}^{(2)}$ & 
$Q_{R}^{(3)}$ & $\sigma$ & $S$  \\ \hline 
$X_g$ & 
$5$ & $-5$ &$3$ & $-4$ & $4$ & 3 &9 & 2   \\
$X_{FN}$ & 
$-5$ & $5$ & $3$ & $4$ & $-4$ & $-1$ & $-9$ & $-4$    \\
$X_{PQ}$ & 
0 & 0 & 6 & 0 & 0 & 2 & 0 & $-2$   \\
$Z_2$ & 
$+$ & $+$ & $+$ & $+$ & $+$ & $+$ & $+$ & $+$   \\ \hline
\end{tabular}
\end{center}

{\footnotesize {\bf Table~1}~~ The $U(1)_g\times U(1)_{FN}$ charge of 
the color triplet fermions $Q_{L,R}^{(i)}$ and the SM singlet complex scalars
$\sigma$ and $S$.
The charge $X_{PQ}$ of $U(1)_{\rm PQ}$ is defined as $X_{PQ}=X_g+X_{FN}$ where
$X_g$ and $X_{FN}$ are charges of $U(1)_g$ and $U(1)_{FN}$, respectively.   
Parity of $Z_2$ which remains after the $U(1)_{PQ}$ breaking by
$\langle S\rangle$ is also listed.}  
\end{figure}

Each degenerate vacuum is separated by potential barriers called domain
wall \cite{dw}. It can be identified with a topological defect
which is produced through the spontaneous breaking of $Z_{|N|}$.
$|N|$ is called domain wall number and it is written as $N_{DW}$
for definiteness in the following part. 
In $N_{DW}=1$ case, the walls are produced although the vacuum is unique. 
They have a string at its boundary which is generated due to
the breaking of $U(1)_{PQ}$. 
This type of domain wall can quickly disappear as studied in \cite{ev}.
On the other hand, in $N_{DW}\ge 2$ case, each string has $N_{DW}$ domain 
walls and they generate complex networks of strings and walls. 
Since these networks are stable, they dominate the energy density of 
the Universe to over-close it. Thus, if inflation does not occur after
the $U(1)_{PQ}$ breaking, the present Universe cannot be realized
unless $N_{DW}=1$ is satisfied.
Inflation could make the present Universe to be covered with
a unique vacuum if inflation occurs after the PQ symmetry breaking.
Thus, low scale inflation could give a solution
in the $N_{DW}\ge 2$ case. However, we focus
on a $N_{DW}=1$ case in the present study.

Here two points on the domain wall problem should be remarked.
First, a non-degenerate vacuum can be realized even for the case
with $N\not=\pm 1$.
As an interesting example, we may consider a $N=\pm 2$ case where the
VEV of the scalar does not break $Z_2$ spontaneously. 
Since two vacua could be identified each other by this unbroken $Z_2$ symmetry,
$N_{DW}=1$ is realized just as in the $N=\pm 1$ case.
Second, we should notice that there are two estimations for axion relic density
by taking account of the decay of domain walls in the
case $N_{DW}=1$ \cite{dwdecay}, which give different conclusions.
One of them suggests that the domain wall problem might not be
solved even in the case $N_{DW}=1$ unless the axion decay constant
is less than a certain limit. Another one claims that the axion
produced through the domain wall decay is subdominant
in comparison with the one due to axion misalignment. 
In the following discussion, we assume that the axion energy
density coming from the domain wall decay is subdominant
and $N_{DW}=1$ could be a solution for the strong $CP$ problem.
 
Now, we try to construct a model so as to escape the domain 
wall problem by $N_{DW}=1$ \cite{s,s1} and to guarantee the goodness of
global symmetry at a required level by gauge symmetry. 
A framework to keep the goodness of the PQ symmetry has 
been proposed in \cite{bs}.
We would like to follow a similar scenario to it.

We impose $U(1)_g\times U(1)_{FN}$ on the model above an intermediate 
scale and introduce new fields with the charge of this symmetry. 
They are two SM singlet complex scalars 
$\sigma, ~S$, and also six types of color triplet fermions 
$Q_{L,R}^{(i)}~(i=1 \sim 3)$, which are assumed
to have no charge of $SU(2)_L\times U(1)_Y$ and 
their subscripts $L$ and $R$ represent their chirality. 
The $U(1)_g\times U(1)_{FN}$ charge of these fields are given in Table 1.
In this charge assignment,  each VEV of 
$\sigma$ and $S$ induces the symmetry breaking
\begin{equation}
U(1)_g\times U(1)_{FN}\quad 
\stackrel{\langle\sigma\rangle}{\longrightarrow}\quad U(1)_{PQ}\quad 
\stackrel{\langle S\rangle}{\longrightarrow}\quad Z_2,  
\label{ssb}
\end{equation}
where we assume $\langle\sigma\rangle >\langle S\rangle$.
The $U(1)_{PQ}$ charge $X_{PQ}$ is defined as a linear combination
$X_{PQ}=X_g+X_{FN}$ where $X_g$ and $X_{FN}$ are the
charges of $U(1)_g$ and $U(1)_{FN}$, respectively. 
As we find it later, $Z_2$ is not be broken through quark condensate either.

We have to address various anomalies associated to the 
introduction of new fields, first of all. 
All of the gauge anomaly for $[SU(3)_c]^3$, $U(1)_g[SU(3)_c]^2$ and
$[U(1)_g]^3$ are easily found to be cancelled within these field contents.  
On the other hand, the QCD anomaly $U(1)_{FN}[SU(3)_c]^2$ for $U(1)_{FN}$
is not cancelled but it is calculated as $N=-2$ in this extra fermion sector.
Since $U(1)_{PQ}$ plays its role as a global symmetry after the first step of the 
symmetry breaking in eq.~(\ref{ssb})
and $U(1)_{PQ}[SU(3)_c]^2$ anomaly takes the same 
value as $U(1)_{FN}[SU(3)_c]^2$,
the strong $CP$ problem is expected to be solved by the PQ mechanism 
based on an axion caused in the spontaneous symmetry breaking of 
$U(1)_{PQ}$ due to a VEV of $S$. 
In order to escape the domain wall problem, the total anomaly including 
contribution from quark sector should be 
$N=\pm 1$ or $\pm 2$.\footnote{If $\langle S\rangle$ and the 
quark condensates do not break a subgroup 
$Z_2$ of $U(1)_{PQ}$, two vacua can be identified by this $Z_2$ 
  symmetry and then the model with $|N|=2$ can be considered to have
  $N_{DW}=1$.}
This suggests that the corresponding anomaly of the quark sector 
should take a value among $0$, $1$, $3$ and $4$.
As we see it later, this value is closely related to the quark mass 
hierarchy and the CKM mixing. 
Three examples (i) $\sim$ (iii) of the charge assignment for the quark sector is 
presented in Table 2. In these cases, $N_{DW}=1$ can be realized.

Next, we move to the problem on the goodness of this $U(1)_{PQ}$ 
and the mass generation of extra colored fermions. 
It is easy to see that a lowest order term in the potential of 
$\sigma$ and $S$, which is $U(1)_g$ invariant but $U(1)_{FN}$
violating, is
\begin{equation}
  \frac{g}{M_{\rm pl}^7}\sigma^{\ast 2} S^9 +{\rm h.c.},
\label{gr-effect}
\end{equation} 
where $g$ is a constant and $U(1)_{FN}$ violation is considered 
to be induced by the 
gravitational effect so that the operator is suppressed by 
Planck mass $M_{\rm pl}$.
If the PQ mechanism works well in this model, the contribution to the 
axion mass from eq.~(\ref{gr-effect}) should be less than the one coming from 
the potential (\ref{axionpot}) due to the QCD instanton effect \cite{bs}.
Since the latter is given as $m_a^2=\frac{m_\pi^2f_\pi^2}{f_a^2}$ \cite{ww},
this condition gives the constraint on $\langle\sigma\rangle$ such as
\begin{equation}
\langle\sigma\rangle~{^<_\sim}~  6\times 10^{12}
\left(\frac{10^{11}~{\rm GeV}}{\langle S\rangle}\right)^{\frac{9}{2}}~{\rm GeV}.  
\label{scale}
\end{equation}
It should be
consistent with our assumption for the symmetry breaking
pattern (\ref{ssb}) within the astrophysical and 
cosmological constraints on the axion decay constant
which is $10^9~{\rm GeV} < f_a <10^{12}~{\rm GeV}$ \cite{strongcp}.
It requires that the VEV of $S$ should satisfy
\begin{equation}
10^9~{\rm GeV}~{^<_\sim}~\langle S\rangle
~{^<_\sim}~2\times 10^{11}~{\rm GeV},
\label{goodness}
\end{equation}
for the $N_{DW}=1$ case.
It suggests that the axion seems to be difficult to be a dominant
component of DM since $f_a$ have to be rather small in this scenario.
From these discussions, we find that the axion in this model is characterized
by a lower mass bound such as $m_a~{^>_\sim}~6\times 10^{-5}$~eV
and a coupling with photon such as
$g_{a\gamma\gamma}=\frac{m_a}{\rm eV}
\frac{2.0}{10^{10}{\rm GeV}}(\frac{E}{N}-1.92)$ \cite{lmn} where
$\frac{E}{N}=-\frac{58}{3}$ for (i), $\frac{34}{3}$ for (ii), and
6 for (iii).

\begin{figure}[t]
\begin{center} 
\begin{tabular}{cc||ccccccccc}\hline
 && $q_{L1}$ & $q_{L2}$ & $q_{L3}$ & $u_{R1}$ & $u_{R2}$ & $u_{R3}$  
&  $d_{R1}$ & $d_{R2}$ & $d_{R3}$  \\ \hline 
(i)~N=1& $X_{FN}$& $-4$ & $-2$ & 0 & 4 &2 & 0 & $-10$ & $-8$ & 2    \\
(ii)~N=4  &  $X_{FN}$&$-8$ & $-2$ & 0 & $-16$ & $-4$ & 0 & $4$ & $2$ & 2    \\
(iii)~N=3  &$X_{FN}$&  $-8$ & $-2$ & 0 & $-16$ &$-4$ & 0 & $2$ & $2$ & 2    \\ 
& $Z_2$ &$+$ & $+$ & $+$ & $+$ & $+$ & $+$ & $+$ & $+$ & $+$  \\
\hline
\end{tabular}
\end{center}

{\footnotesize {\bf Table~2}~~ Examples of the $U(1)_{FN}$ charge 
assignment for quarks which have no domain wall problem.
Quarks are assumed to have no $U(1)_g$ charge.
$N$ represents a value of the QCD anomaly in each case, which realizes 
$N_{DW}=1$ by combining with the one from the extra colored fermions
given in Table 1. }  
\end{figure}  

The extra colored fields can get their mass only through the VEVs of 
$\sigma$ and $S$. It is crucial for the consistency of the model
what scale of masses they can have.
The following operators are invariant under 
$U(1)_g\times U(1)_{FN}$, 
\begin{equation}
\sigma \bar Q_{L}^{(1)}Q_{R}^{(1)}, \qquad \sigma\bar Q_{L}^{(2)}Q_{R}^{(2)}.
\end{equation}
On the other hand, $\frac{S^{\ast 2}}{M_\ast}\bar Q_L^{(3)}Q_R^{(3)}$
could be generated as a $U(1)_{PQ}$ invariant operator after the 
$U(1)_g\times U(1)_{FN}$ breaking at $\langle \sigma\rangle=M_\ast$. 
These operators give masses to these extra colored fermions through
$\langle\sigma\rangle$ and $\langle S\rangle$.
However, since they have no hypercharge, they cannot couple with ordinary 
quarks and then have no decay modes to be stable.\footnote{It may be
possible to assume that these fermions have hypercharge and couple with
ordinary quarks to have decay modes. However, in that case, 
we have to introduce a lot of fields to cancel the gauge anomaly. 
We do not consider such a possibility here.}
If they are in thermal equilibrium during the history of the Universe, 
we have to note that several contradictions such as the existence
of fractionally charged hadrons and their over-abundant
contribution to the energy density could appear \cite{lmn}.
The most strong constraint on their abundance comes from searches of
fractionally charged particles, which requires
$\frac{n_{Q^{(3)}}}{n_b}~{^<_\sim}~10^{-20}$ for the abundance of $Q^{(3)}$ and
ordinary nucleons \cite{fract}.
This constraint could be satisfied even if $Q^{(3)}$ is in the thermal
equilibrium, as long as reheating temperature is assumed to be much lower than 
the mass of $Q^{(3)}$ which is the lightest extra colored fermion.
Since $U(1)_g\times U(1)_{FN}$ is assumed not to be restored after
the reheating, these extra colored fermions are not produced
in the thermal bath through the reheating process and the model
can escape this problem.
In fact, we can confirm that the $Q^{(3)}$ mass $O\left(\frac{\langle S\rangle}{M_\ast}\right)^2M_\ast$ derived by an $O(1)$ coupling could satisfy the above constraint for parameters used in the following study and the 
reheating temperature such as $T_R=10^8$~GeV. 
Such a low reheating temperature could cause a problem if we consider
the thermal leptogenesis due to the decay of thermal right-handed
neutrinos. We will come back this point later.   
 
Now, we couple this model with the SM including a lepton sector.
Since the axion could not be a dominant component of DM in this scenario
as discussed above, we need to prepare a candidate for DM.
For this purpose, the leptonic sector is extended by an additional
doublet scalar $\eta$ and three right-handed neutrinos $N_i$ so as to
realize the scotogenic model \cite{ma,radnm,ks}.
An example of the $U(1)_g\times U(1)_{FN}$ charge assignment
for the leptonic sector is shown in Table 3.
After the symmetry breaking due to 
$\langle\sigma\rangle$, $U(1)_{PQ}$ invariant operators are considered to
be generated in both Yukawa couplings and a scalar potential 
of an effective theory at energy regions below  $\langle\sigma\rangle$.
An interesting point is that nonrenormalizable Yukawa couplings
are controlled by the $U(1)_{PQ}$ charge of each quark and lepton
\cite{fn-pq,s1}.
In fact, if we define
\begin{eqnarray}
&&n^u_{ij}=\frac{1}{2}(X_{u_{R_j}}-X_{q_{L_i}}), \quad 
n^d_{ij}=\frac{1}{2}(X_{d_{R_j}}-X_{q_{L_i}}),  \quad
n^N_{ij}=\frac{1}{2}(X_{N_{R_i}}+X_{N_{R_j}}), \nonumber \\
&&n^\nu_{ij}=\frac{1}{2}(X_{N_{R_j}}-X_{\ell_{L_i}}-1), \quad 
n^e_{ij}=\frac{1}{2}(X_{e_{R_j}}-X_{\ell_{L_i}}), 
\label{order}
\end{eqnarray}
quark Yukawa couplings are written as
\begin{eqnarray}
-{\cal L}_y^q&=&\sum_{i=1,j}^3\left[y^u_{ij}\left(\frac{S}{M_\ast}
\right)^{|n^u_{ij}|}\bar q_{L_i}\phi u_{R_j}
+y^d_{ij}\left(\frac{S}{M_\ast}\right)^{|n^d_{ij}|}
\bar q_{L_i}\tilde\phi d_{R_j} \right], 
\label{yukawa0}
\end{eqnarray} 
where $\tilde\phi=i\tau_2\phi^\ast$ and $M_\ast=\langle\sigma\rangle$.
On the other hand, Yukawa couplings relevant to neutrino mass generation 
are written as
\begin{eqnarray}
-{\cal L}_y^\ell = \sum_{i,j=1}^3\left[h_{ij}^\nu
\left(\frac{S}{M_\ast}\right)^{|n^\nu_{ij}|}\bar \ell_{L_i}\eta N_{R_j} \right.
&+&h_{ij}^e\left(\frac{S}{M_\ast}\right)^{|n^e_{ij}|}
\bar \ell_{L_i}\tilde\phi e_{R_j}  \nonumber\\ 
&+& \left. h_{ij}^N\left(\frac{S}{M_\ast}\right)^{|n_{ij}^N|}
M_\ast\bar N^c_{R_i}N_{R_j} +{\rm h.c.}\right]. 
\label{yneut} 
\end{eqnarray}
The third term related to the mass of right-handed neutrinos should satisfy
$|n_{ij}^N|\ge 2$, since renormalizable one is forbidden by
$U(1)_g\times U(1)_{FN}$. 
In these formulas (\ref{yukawa0}) and (\ref{yneut}),
$S$ should be replaced by $S^\ast$ for $n^f_{ij}<0$.
The scalar potential at energy regions lower than $\langle \sigma\rangle$
is written as
\begin{eqnarray}
V_1&=&m_S^2S^\dagger S+\kappa_1(S^\dagger S)^2+\kappa_2(S^\dagger S)(\phi^\dagger\phi)
+\kappa_3(S^\dagger S)(\eta^\dagger\eta) \nonumber \\
&+&m_\eta^2\eta^\dagger\eta +m_\phi^2\phi^\dagger\phi
+\lambda_1(\phi^\dagger\phi)^2
+\lambda_2(\eta^\dagger\eta)^2 
+\lambda_3(\phi^\dagger\phi)(\eta^\dagger\eta) 
+\lambda_4(\phi^\dagger\eta)(\eta^\dagger\phi) \nonumber \\ 
&+&\frac{\lambda_5}{2}\left[\frac{S}{M_\ast}(\eta^\dagger\phi)^2
~+{\rm h.c.}\right],
\label{smodel}
\end{eqnarray}
where $\lambda_5$ is taken to be real. On the other hand, the scalar
potential for the light scalars $\phi$ and $\eta$ after $S$ gets the VEV
can be expressed as
\begin{eqnarray}
V_2&=&\tilde m_\eta^2\eta^\dagger\eta +\tilde m_\phi^2\phi^\dagger\phi
+\tilde\lambda_1(\phi^\dagger\phi)^2
+\tilde\lambda_2(\eta^\dagger\eta)^2 \nonumber \\ 
&+&\tilde\lambda_3(\phi^\dagger\phi)(\eta^\dagger\eta) 
+\lambda_4(\phi^\dagger\eta)(\eta^\dagger\phi) 
+\frac{\tilde\lambda_5}{2}\left[(\eta^\dagger\phi)^2
~+{\rm h.c.}\right],
\label{scot}
\end{eqnarray}
which is found to coincide with the scalar potential of the scotogenic model. 

\begin{figure}[t]
\begin{center} 
\begin{tabular}{c||ccccccccccc}\hline
 & $\ell_{L1}$ & $\ell_{L2}$ & $\ell_{L3}$ & $e_{R1}$ & $e_{R2}$ & $e_{R3}$  
 &  $N_{R1}$ & $N_{R2}$ & $N_{R3}$  & $\phi$ & $\eta$ \\ \hline 
$X_{FN}$ &  $-6$ & $-2$ & $-2$ & 4 &2 &2 & $3$ & $1$ & $-1$ & 0 & $-1$   \\
$Z_2$ & $+$ & $+$ & $+$ & $+$ & $+$ & $+$ & $-$ & $-$ &
 $-$ &  $+$ & $-$   \\ \hline
\end{tabular}
\end{center}

{\footnotesize {\bf Table~3}~~ The $U(1)_{FN}$ charge assignment
for leptons, right-handed neutrinos, the Higgs doublet $\phi$ and 
an additional doublet scalar $\eta$. These are assumed to have 
no $U(1)_g$ charge.}  
\end{figure}

In eqs.~(\ref{smodel}) and (\ref{scot}),
scalar masses and couplings are shifted from ones
at higher energy regions due to the symmetry breaking effect
by $\sigma$ and $S$, respectively \cite{s}.
The shift of parameters in (\ref{smodel}) can be summarized as 
\begin{eqnarray}
&&\kappa_1=\bar\kappa_1-\frac{\xi_S^2}{4\xi_\sigma}, \quad
\kappa_2=\bar\kappa_2-\frac{\xi_S\xi_\phi}{2\xi_\sigma}, \quad
\kappa_3=\bar\kappa_3-\frac{\xi_S\xi_\eta}{2\xi_\sigma}, \nonumber\\
&&\lambda_1=\bar\lambda_1-\frac{\xi_\phi^2}{4\xi_\sigma}, \quad
\lambda_2=\bar\lambda_2-\frac{\xi_\eta^2}{4\xi_\sigma}, \quad
\lambda_3=\bar\lambda_3-\frac{\xi_\phi\xi_\eta}{2\xi_\sigma}, \nonumber\\
&&m_S^2=\bar m_S^2+\xi_S\langle \sigma\rangle^2, \quad
m_\phi^2=\bar m_\phi^2+\xi_\phi\langle \sigma\rangle^2, \quad
m_\eta^2=\bar m_\eta^2+\xi_\eta\langle \sigma\rangle^2,
\label{gcoupl}
\end{eqnarray}
where over-lined parameters correspond to the ones before the
symmetry breaking and $\xi_\rho~(\rho=\sigma,S,\phi,\eta)$ represents a
coupling constant for an operator $(\rho^\dagger\rho)(\sigma^\dagger\sigma)$
in the potential at energy scales
larger than $\langle\sigma\rangle$.

On the other hand, the shift of parameters in (\ref{scot})
can be given as 
\begin{eqnarray}
&&\tilde\lambda_1=\lambda_1-\frac{\kappa_2^2}{4\kappa_1}, \quad
\tilde\lambda_2=\lambda_2-\frac{\kappa_3^2}{4\kappa_1}, \quad
\tilde\lambda_3=\lambda_3-\frac{\kappa_2\kappa_3}{2\kappa_1}, \nonumber\\
&&\tilde\lambda_5=\lambda_5\frac{\langle S\rangle}{M_\ast}, \quad
\tilde m_\phi^2= m_\phi^2+\kappa_2\langle S\rangle^2, \quad
\tilde m_\eta^2= m_\eta^2+\kappa_3\langle S\rangle^2.
\label{gcoupl1}
\end{eqnarray}
The parameters in eq.~(\ref{gcoupl1}) should satisfy conditions
for a vacuum defined in $V_2$ to be stable, which are written as 
\begin{equation}
  \tilde\lambda_{1,2}>0,\quad
  \tilde\lambda_3>-2\sqrt{\tilde\lambda_1\tilde\lambda_2},\quad
  \tilde\lambda_3+\lambda_4-|\tilde\lambda_5| >-2
  \sqrt{\tilde\lambda_1\tilde\lambda_2}.
  \label{stab}
\end{equation}  

In these equations, the lowest dimension operators invariant under $U(1)_{PQ}$
are listed for nonrenormalizable ones.  
There could be $U(1)_{FN}$ violating contributions to them which are
induced by the gravity effect. However, since they are suppressed by a factor
$\frac{\sigma\sigma^\ast}{M_{\rm pl}^2}$ at least,
their effect can be safely neglected under the condition (\ref{scale}).
These formulas show that Yukawa couplings for the quarks and the leptons have
a suppression by powers of $\frac{|\langle S\rangle|}{M_\ast}$ after
the PQ symmetry breaking due to $\langle S\rangle$. 
Neutrino Yukawa couplings in the leptonic sector are also found to reduce
to the ones in the scotogenic model.
Moreover, $\tilde\lambda_5$ term in eq.~(\ref{scot}) could be
small so as to cause small mass difference between neutral
components of the extra doublet scalar $\eta$.
One should remind that it is a crucial element of the 
neutrino mass generation in the original
scotogenic model.

\section{Phenomenological features of the model}
\subsection{Quark mass hierarchy and CKM mixing}
After the PQ symmetry breaking due to $\langle S\rangle$, 
 eq.~(\ref{yukawa0})
induces Yukawa couplings for quarks with a suppression factor
$\epsilon^{|n^f_{ij}|}$ where $\epsilon=\frac{|\langle S\rangle|}{M_\ast}$ 
and $n^f_{ij}$ is determined by the PQ charge of quarks just like
Froggatt-Nielsen mechanism \cite{fn-pq,s1}.\footnote{
In the different context, flavor structure of quarks and leptons have been
extensively studied using flavons resulting from various types of 
flavor symmetry \cite{wil, flav-pq,flavon-fcnc}.}
Elements of quark mass matrices derived from these are represented as
\begin{equation}
m^f_{ij}=y^f_{ij}\epsilon^{|n^f_{ij}|}\langle \phi\rangle,
\label{m-eigen}
\end{equation}
where a superscript $f$ stands for up and down sector and then $f=u, d$. 
If we define the quark mass eigenstates as 
$\tilde f_L=U^ff_L$ and $\tilde f_R=V^ff_R$ using
the unitary matrices $U^f$ and $V^f$, they satisfy the condition  
\begin{equation}
\left(U^{f\dagger}\right)_{\alpha i}y^f_{ij}\epsilon^{|n^f_{ij}|}V^{f}_{j\beta}
=\frac{m_\alpha^f}{\langle \phi\rangle}~\delta_{\alpha\beta},
\label{c-diag}
\end{equation}
where $m^f_\alpha$ represents a mass eigenvalue in the $f$-sector.
The CKM matrix is expressed as $U_{CKM}=U^{u\dagger}U^d$.
If we use the PQ charge of quarks given in Table~2, 
the quark mass matrices defined by $\bar u_L{\cal M}_uu_R$ and 
$\bar d_L{\cal M}_dd_R$ can be written for each example as
\begin{eqnarray}
&{\rm (i)}&{\cal M}_u=\left(\begin{array}{ccc}
y_{11}^u~\epsilon^4 & y_{12}^u~\epsilon^3 & y_{13}^u~\epsilon^2 \\
y_{21}^u~\epsilon^3 & y_{22}^u~\epsilon^2 & y_{23}^u~\epsilon \\
y_{31}^u~\epsilon^2 & y_{32}^u~\epsilon & y_{33}^u \\
\end{array}\right)\langle \phi\rangle , \quad
{\cal M}_d=\left(\begin{array}{ccc}
y_{11}^d~\epsilon^3 & y_{12}^d~\epsilon^2 & y_{13}^d~\epsilon^3 \\
y_{21}^d~\epsilon^4 & y_{22}^d~\epsilon^3 & y_{23}^d~\epsilon^2 \\
y_{31}^d~\epsilon^5 & y_{32}^d~\epsilon^4 & y_{33}^d~\epsilon \\
\end{array}\right)\langle \tilde\phi\rangle ,  \nonumber \\
&{\rm (ii)}&{\cal M}_u=\left(\begin{array}{ccc}
y_{11}^u~\epsilon^4 & y_{12}^u~\epsilon^2 & y_{13}^u~\epsilon^4 \\
y_{21}^u~\epsilon^7 & y_{22}^u~\epsilon & y_{23}^u~\epsilon \\
y_{31}^u~\epsilon^8 & y_{32}^u~\epsilon^2 & y_{33}^u \\
\end{array}\right)\langle \phi\rangle , \quad
{\cal M}_d=\left(\begin{array}{ccc}
y_{11}^d~\epsilon^6 & y_{12}^d~\epsilon^5 & y_{13}^d~\epsilon^5 \\
y_{21}^d~\epsilon^3 & y_{22}^d~\epsilon^2 & y_{23}^d~\epsilon^2 \\
y_{31}^d~\epsilon^2 & y_{32}^d~\epsilon & y_{33}^d~\epsilon \\
\end{array}\right)\langle \tilde\phi\rangle , \nonumber \\
&{\rm (iii)}&{\cal M}_u=\left(\begin{array}{ccc}
y_{11}^u~\epsilon^4 & y_{12}^u~\epsilon^2 & y_{13}^u~\epsilon^4 \\
y_{21}^u~\epsilon^7 & y_{22}^u~\epsilon & y_{23}^u~\epsilon \\
y_{31}^u~\epsilon^8 & y_{32}^u~\epsilon^2 & y_{33}^u \\
\end{array}\right)\langle \phi\rangle , \quad
{\cal M}_d=\left(\begin{array}{ccc}
y_{11}^d~\epsilon^5 & y_{12}^d~\epsilon^5 & y_{13}^d~\epsilon^5 \\
y_{21}^d~\epsilon^2 & y_{22}^d~\epsilon^2 & y_{23}^d~\epsilon^2 \\
y_{31}^d~\epsilon & y_{32}^d~\epsilon & y_{33}^d~\epsilon \\
\end{array}\right)\langle \tilde\phi\rangle , \nonumber\\
\end{eqnarray}

While flavor dependent PQ charge of quarks could bring about these 
mass matrices, it can also cause flavor changing neutral processes
with axion emission \cite{wil,fn-pq}, which can be severely constrained through
experiments. The strongest constraint on $f_a$ due to such processes is
known to come from $K^\pm \rightarrow \pi^\pm a$, 
whose experimental bound is given as
${\rm Br}(K^\pm \rightarrow \pi^\pm a)< 7.3\times 10^{-11}$ \cite{exkpa}. 
Since the axion $a$ is introduced in the effective theory
through the replacement $S=\langle S\rangle e^{i\frac{a}{f_a}}$,
eq.~(\ref{yukawa0}) gives axion-quark interaction terms
\begin{equation}
in^u_{ij}m^u_{ij}~\frac{a}{f_a}~\bar u_{Li}u_{Rj}+
in^d_{ij}m^d_{ij}~\frac{a}{f_a}~\bar d_{Li}d_{Rj}+{\rm h.c.},
\label{a-int}
\end{equation}
where $m^f_{ij}$ is given in eq.~(\ref{m-eigen}).
If we focus our attention to the down-sector and use the quark 
mass eigenstates defined above, corresponding terms in
eq.~(\ref{a-int}) can be rewritten as
\begin{eqnarray}
  &&i\frac{\langle \phi\rangle}{f_a}\left[\left(U^{d\dagger}\right)_{\alpha i}
 n^u_{ij}y^d_{ij}\epsilon^{n^u_{ij}}V^{d}_{j\beta}-
\left(V^{d\dagger}\right)_{\alpha i}n^u_{ij}y^{\ast d}_{ji}\epsilon^{n^u_{ij}}
U^{d}_{i\beta}\right]
a~\bar d_\alpha d_\beta \nonumber \\
&+&i\frac{\langle \phi\rangle}{f_a}\left[\left(U^{d\dagger}\right)_{\alpha i}
  n^u_{ij}y^d_{ij}\epsilon^{n^u_{ij}}V^{d}_{j\beta}+
\left(V^{d\dagger}\right)_{\alpha i}n^u_{ij}y^{\ast d}_{ji}
\epsilon^{n^u_{ij}}U^{d}_{i\beta}\right]
a~\bar d_\alpha\gamma_5 d_{\beta} \nonumber \\
&\equiv&  iS_{\alpha\beta}~a~\bar d_\alpha d_\beta 
+ iA_{\alpha\beta}~a~\bar d_\alpha\gamma_5 d_\beta  . 
\label{a-int-m}
\end{eqnarray}
If we apply eqs.~(\ref{order}) and (\ref{c-diag}) to eq.~(\ref{a-int-m}),
the coupling constants $S_{\alpha\beta}$ and $A_{\alpha\beta}$
are found to be expressed as
\begin{equation}
S_{\alpha\beta}=\frac{m_\alpha-m_\beta}{2f_a}X_{\alpha\beta}^+, \qquad 
A_{\alpha\beta}=\frac{m_\alpha+m_\beta}{2f_a}X_{\alpha\beta}^-, \nonumber \\
\end{equation}
where $X_{\alpha\beta}^\pm$ is defined by
\begin{equation}
X_{\alpha\beta}^\pm=
\left(V^{d\dagger}\right)_{\alpha i}X(d_{Ri})\left(V^{d}\right)_{i\beta}
\pm\left(U^{d\dagger}\right)_{\alpha i}X(d_{Li})\left(U^{d}\right)_{i\beta}.
\end{equation}
Since the decay width of $K^+ \rightarrow \pi^+ a$ can be estimated by using
this $X_{\alpha\beta}^\pm$ as \cite{fn-pq,bfsym}
\begin{equation}
  \Gamma=\frac{|X_{ds}^+|^2}{128\pi}\frac{m_K^3}{f_a^2}
  \left(1-\frac{m_\pi^2}{m_K^2}\right)^3,
\end{equation}
 we obtain 
the strong constrain on $f_a$ by applying the experimental bound
to this formula as 
\begin{equation}
f_a> 2.4\times 10^{11}~ |X_{ds}^+|~ {\rm GeV}.
\label{fcn-a}
\end{equation}
On the other hand, since the condition (\ref{goodness}) should be satisfied,
eq.~(\ref{fcn-a}) requires $|X_{ds}^+|<1$. 
The PQ charge of quarks is required not only to
reproduce the quark mass eigenvalues and the CKM mixing
but also to satisfy this constraint. 

We examine these issues in the examples shown in Table 2.
Since these realize $N_{DM}=1$, the axion decay constant $f_a$
satisfies $f_a=|\langle S\rangle|$.
In order to study features of the examples quantitatively, we need to fix a
value of $\epsilon$ and coupling constants $y_{ij}^f$.
Needless to say, the validity of the scenario is determined through
how good predictions can be derived for 
less number of independent coupling constants $y_{ij}^f$ without serious 
fine tuning.
The results in each example are ordered for a typical parameter set.
In this analysis, the $CP$ phase of $y_{ij}^f$ is not taken into account, 
for simplicity.

In example (i), we assume $\epsilon=0.08$ and
the coupling constants $y_{ij}^f$ are fixed as 
\begin{eqnarray}
&&y_{11}^u=y_{23}^u=y_{32}^u=y_{33}^u=1,\quad
y_{13}^u=y_{22}^u=y_{31}^u=0.1, \quad
y_{12}^u=y_{21}^u=0.7,   \nonumber  \\
&&y_{21}^d=y_{22}^d=y_{31}^d=y_{32}^d=1, \quad y_{11}^d=y_{13}^d=y_{23}^d=0.1,
\quad y_{12}^d=0.022, \quad  y_{33}^d=0.3,  \nonumber
\end{eqnarray}
where the number of independent parameters can be identified as six.
For this parameter set, the quark mass eigenvalues and the CKM matrix
are obtained as
{\footnotesize
\begin{equation}
\begin{array}{lll}
m_u\simeq 2.6~{\rm MeV}, & m_c\simeq 1.1~{\rm GeV}, & 
m_t\simeq 174~{\rm GeV}, \\ 
m_d\simeq 6.7~{\rm MeV}, & m_s\simeq 92~{\rm MeV}, & 
m_b\simeq 4.2~{\rm GeV}, 
\end{array}
\quad
V_{\rm CKM}\simeq\left(\begin{array}{ccc}
0.97 &-0.23 & -0.0052 \\
0.23 & 0.97 & -0.018 \\
0.0092 & 0.017 & 1.0 \\
\end{array}\right).
\nonumber
\end{equation}
}
In this case, eq.~(\ref{fcn-a}) requires $f_a> 1.7\times 10^{11}$~GeV.

In example (ii), $\epsilon=0.07$ is assumed and $y_{ij}^f$ are fixed as 
\begin{eqnarray}
&&y_{11}^u=y_{13}^u=y_{21}^u=y_{31}^u=y_{32}^u=y_{33}^u=1,\quad
y_{22}^u=y_{23}^u=0.1, \quad
y_{12}^u=0.32,   \nonumber  \\
&&y_{11}^d=y_{21}^d=y_{31}^d=1, \quad y_{22}^d=0.1, \quad y_{23}^d=-0.03,
\quad y_{32}^d= y_{33}^d=0.26.\quad y_{12}^d= y_{13}^d=60, \nonumber 
\end{eqnarray}
where the number of independent parameters can be identified as seven.
For this parameter set, we obtain 
{\footnotesize
\begin{equation}
\begin{array}{lll}
m_u\simeq 4.0~{\rm MeV}, & m_c\simeq 1.3~{\rm GeV}, & 
m_t\simeq 174~{\rm GeV}, \\ 
m_d\simeq 3.9~{\rm MeV}, & m_s \simeq 93~{\rm MeV}, & 
m_b\simeq 4.6~{\rm GeV}, 
\end{array}
\quad
V_{\rm CKM}\simeq \left(\begin{array}{ccc}
0.97 &0.24 & 0.0042 \\
-0.24 & 0.97 & -0.0056 \\
-0.0054 & 0.0043 & 1.0 \\
\end{array}\right).
\nonumber
\end{equation}
}
Eq.~(\ref{fcn-a}) requires $f_a> 2.2\times 10^{11}$~GeV.

In example (iii), if we assume the same values for $\epsilon$
and $y^u_{ij}$ as the ones in the example (ii), and $y^d_{ij}$ are taken as 
\begin{eqnarray}
&&y_{11}^d=y_{21}^d=y_{31}^d=\epsilon, \quad y_{22}^d=0.1, \quad y_{23}^d=-0.03,
\quad y_{32}^d= y_{33}^d=0.26.\quad y_{12}^d= y_{13}^d=60. \nonumber
\end{eqnarray}
Since ${\cal M}_u$ and ${\cal M}_d$ take the same form as the ones of
the example (ii),  the quark
mass eigenvalues and the CKM matrix take the same values as the ones
in the example (ii).
The number of independent parameters used here can be identified
as eight. The bound on $f_a$ is estimated as $f_a> 1.3\times 10^{10}$~GeV,
which is one order of magnitude smaller than the previous two examples.

These examples show that the constraint on $f_a$ coming from the flavor
dependent PQ charge assignment could be much stronger than the
astrophysical constraint as suggested in \cite{fn-pq}.  However, it could be
consistent with the cosmological upper bound of $f_a$ even if
the realization of realistic values for the quark mass eigenvalues and
the CKM mixing is imposed.
On the other hand, the consistency of this constraint with
the upper bound of $f_a$ imposed by the goodness of the PQ symmetry
could depend largely on the PQ charge assignment.
In fact, although the consistency is complete in the example (iii),
the situation is marginal in the examples (i) and (ii).
In the example (iii), the scenario is found to work well even if serious
fine tuning of the coupling constants $y_{ij}^f$ is not adopted.
The obtained results seem to be rather good compared with the data
listed in \cite{pdg} although the number of independent parameters are
smaller than the number of physical observables in the quark sector.

\subsection{Leptonic sector}
In this model, the neutrino mass generation is forbidden
at tree-level due to
$U(1)_{PQ}$ even after the breaking of $U(1)_{g}\times U(1)_{FN}$,
since $\eta$ is assumed to have no VEV,
However, since both right-handed neutrino masses and mass 
difference between the neutral components of $\eta$
are induced after the breaking of
$U(1)_{PQ}$ as found form eqs.~(\ref{yneut}) and (\ref{scot}),
small neutrino masses are generated
radiatively in the same way as the original scotogenic model
through a one-loop diagram which is shown in Fig.~1.
If we apply the PQ charge given in Table 3 to eq.~(\ref{yneut}),
the Dirac mass matrix for charged leptons which is
defined by $\bar e_L{\cal M}_e e_R$ and 
the the Majorana mass matrix ${\cal M}_N$ 
for right-handed neutrinos $N_{R_i}$ are
expressed as  
\begin{equation}
{\cal M}_e=\left(\begin{array}{ccc}
h_{11}^e~\epsilon^5 & h_{12}^e~\epsilon^4 & h_{13}^e~\epsilon^4 \\
h_{21}^e~\epsilon^3 & h_{22}^e~\epsilon^2 & h_{23}^e~\epsilon^2 \\
h_{31}^e~\epsilon^3 & h_{32}^e~\epsilon^2 & h_{33}^e~\epsilon^2 \\
\end{array}\right)\langle \tilde\phi\rangle ,  \quad 
{\cal M}_N=\left(\begin{array}{ccc}
h_{11}^N~\epsilon^3 & h_{12}^N\epsilon^2 & h_{13}^N~\epsilon^3 \\
h_{12}^N\epsilon^2 & h_{22}^N~\epsilon^3 & h_{23}^N~\epsilon^2 \\
h_{13}^N~\epsilon^3 & h_{23}^N~\epsilon^2 & h_{33}^N\epsilon^3 \\
\end{array}\right)M_\ast.
\label{r-mass}
\end{equation}
In the mass matrix ${\cal M}_N$, we take account that allowed operators 
start from the nonrenormalizable ones.

\input epsf
\begin{figure}[t]
\begin{center}
\epsfxsize=7.5cm
\leavevmode
\epsfbox{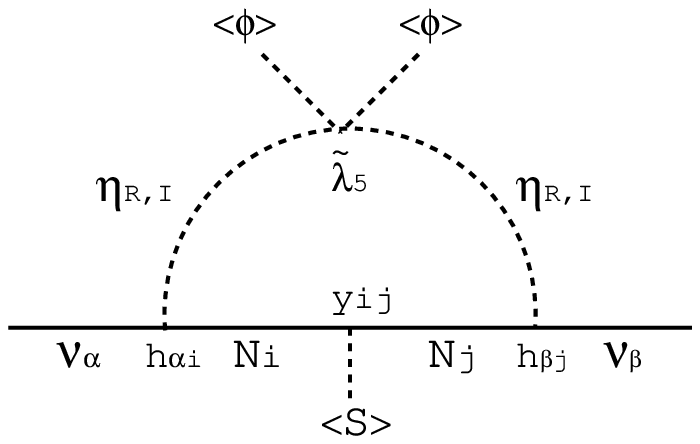}
\end{center}
\vspace*{-3mm}

        {\footnotesize {\bf Fig.~1}~~One-loop diagram for neutrino mass
          generation, in which $\eta_R$ and $\eta_I$ are a real and an imaginary
          part of the neutral component of $\eta$, respectively.}
\end{figure}

This right-handed neutrino mass matrix ${\cal M}_N$ 
suggests that three mass
eigenvalues tend to take the same order values. If we assume values of the
Yukawa coupling constants $h_{ij}^N$ appropriately, 
the eigenvalues of ${\cal M}_N$
can be fixed, for example, as\footnote{In this choice,
  we refer to the previous work \cite{s}.}
\begin{equation}
  M_1\simeq1.0\times 10^8~{\rm GeV}, \qquad
  M_2\simeq 4.2\times 10^8~{\rm GeV}, 
\qquad M_3\simeq 1.9\times 10^9~{\rm GeV},  
\label{rmass}
\end{equation}
where we assume $M_\ast=10^{12}$~GeV.
The neutrino mass generated through a
one-loop diagram can be approximately written as
\begin{equation}
  ({\cal M}_\nu)_{ij}=\sum_{k=1}^3 \tilde h_{ik}^\nu\tilde h_{j k}^\nu\Lambda_k,
  \qquad
\Lambda_k\simeq \frac{\tilde\lambda_5\langle\phi\rangle^2}{8\pi^2M_k}
\ln\frac{M_k^2}{\bar M_\eta^2},
\label{lnmass}
\end{equation} 
where we use $M_{\eta_{R,I}}^2\gg |M_{\eta_R}^2-M_{\eta_I}^2|$, which is
noted in the previous part. 
$M_k$ is a mass eigenvalue of the right-handed neutrino and 
$\bar M_\eta^2=\tilde m_\eta^2+\left(\tilde\lambda_3+\lambda_4\right)
\langle\phi\rangle^2$.
In this formula, $\tilde h_{ij}^\nu$ and $\tilde\lambda_5$ are defined by
using $\epsilon$ as
$\tilde h_{ij}^\nu=h_{ij}^\nu\epsilon^{|n_{ij}^\nu|}$ and
$\tilde\lambda_5=\lambda_5\epsilon$.

Here, it may be useful to note that the present ${\cal M}_\nu$ has
interesting flavor structure consistent with tri-bimaximal mixing
if ${\cal M}_N$ is diagonal.
In fact, if effective neutrino Yukawa coupling constants defined above
satisfy the relation  
\begin{equation}
  \tilde h_{1j}^\nu=0, \quad \tilde h_{2j}^\nu= \tilde h_{3j}^\nu\equiv
  h_j \quad (j=1,2); \qquad 
\tilde h_{13}^\nu=\tilde h_{23}^\nu=-\tilde h_{33}^\nu\equiv h_3, 
\label{flavor}
\end{equation}
${\cal M}_\nu$ is found to be diagonalized by a
tri-bimaximal NMS matrix \cite{tribi} and
mass eigenvalues are derived as
\begin{eqnarray}
&&m_{\nu_1}=0, \qquad m_{\nu_2}= 3|h_3|^2\Lambda_3, \nonumber \\
&&m_{\nu_3}=2\left[|h_1|^4\Lambda_1^2+|h_2|^4\Lambda_2^2+
2|h_1|^2|h_2|^2\Lambda_1\Lambda_2\cos 2(\theta_1-\theta_2)
\right]^{1/2}, 
\label{nmass}
\end{eqnarray}
where $\theta_j={\rm arg}(h_j)$.
On the other hand, if we notice that
neutrino Yukawa interaction in eq.~(\ref{yneut}) takes the form
\begin{equation}
\eta^0~(\bar\nu_{L1}, \bar\nu_{L2}, \nu_{L3})\left(\begin{array}{ccc}
h_{11}^\nu~\epsilon^4 & h_{12}^\nu~\epsilon^3 & h_{13}^\nu~\epsilon^2 \\
h_{21}^\nu~\epsilon^2 & h_{22}^\nu~\epsilon & h_{23}^\nu~\epsilon^2 \\
h_{31}^\nu~\epsilon^2 & h_{32}^\nu~\epsilon & h_{33}^\nu~\epsilon^2 \\
  \end{array}\right)
\left( \begin{array}{c} N_{R1}\\ N_{R2} \\ N_{R3}\\ \end{array}\right),
\label{dracn}
\end{equation}
the above relation (\ref{flavor}) among $\tilde h^\nu_{ij}$ is found to be
realized just by assuming the same relation for $h_{ij}^\nu$ without
changing the suppression structure due to $\epsilon$.
The present PQ charge assignment is consistent with the tri-bimaximal
flavor structure approximately.
However, unfortunately, the present right-handed neutrino mass matrix
${\cal M}_N$ is not diagonal. Although this flavor structure is lost
after ${\cal M}_N$ is diagonalized, this knowledge can be useful to find
suitable neutrino Yukawa couplings $h_{ij}^\nu$ referring the previous
study in \cite{ks}.

In order to see the resulting flavor structure in the leptonic
sector, we take $\epsilon=0.07$, $\tilde m_\eta=1$~TeV, and
$\tilde\lambda_5=5.4\times 10^{-3}$ which corresponds to
$\lambda_5\simeq 0.08$.
The charged lepton coupling constants $h_{ij}^e$ and neutrino
Yukawa coupling constants $h_{ij}^\nu$ are fixed as 
\begin{eqnarray}
&&h_{11}^e=h_{21}^e=h_{31}^e=1,\quad
  h_{32}^e=h_{33}^e=1.47, \quad h_{12}^e=0.82,\quad h_{22}^e=0.17, \nonumber \\
  &&h_{13}^e=0.4,\quad h_{23}^e=0.02, \nonumber\\
  &&h_{11}^\nu=h_{12}^\nu=1, \quad h_{13}^\nu=0.6, \quad
   h_{21}^\nu=h_{31}^\nu=6.5\times 10^{-3}, \quad h_{22}^\nu =0.23, \nonumber \\
   && h_{32}^\nu=0.184, \quad h_{23}^\nu=0, \quad h_{33}^\nu=1.43.
   \label{lyuk}
\end{eqnarray}
For this parameter set, we can obtain 
{\footnotesize
\begin{equation}
\begin{array}{lll}
m_{\nu_1}\simeq 0~{\rm MeV}, & m_{\nu_2}\simeq 8.5\times 10^{-3}~{\rm eV}, & 
m_{\nu_3}\simeq 5.2\times 10^{-2}~{\rm eV}, \\ 
m_e=0.51~{\rm MeV}, & m_\mu=106~{\rm MeV}, & 
m_\tau=1.78~{\rm GeV}, 
\end{array}
\quad
V_{\rm MNS}\simeq \left(\begin{array}{ccc}
0.87 &-0.46 & -0.14 \\
-0.29 & 0.74 & 0.60 \\
0.38 & -0.48 & 0.78 \\
\end{array}\right).
\nonumber
\end{equation}
}
Squared mass differences required by the 
neutrino oscillation data could be explained by these values. 
The NMS matrix is shifted from the tri-bimaximal mixing and
$U_{e3}$ takes a favorable value. 
Although Yukawa coupling constants have to be tuned within the similar order,
the required tuning is not serious one. The suppression due to the PQ symmetry
can be considered to work rather well in the leptonic sector also.

Here, we should comment on a reason why $\tilde h_{i1}^\nu$ is fixed
at small values of $O(10^{-4})$.
It is not for the neutrino mass generation but for
thermal leptogeesis \cite{leptg}.
As is known generally and found also from the present 
neutrino mass formula (\ref{nmass}), 
the neutrino masses required by the neutrino oscillation data 
could be derived only by two right-handed neutrinos.
It means that the mass and the neutrino Yukawa couplings of a remaining 
right-handed neutrino could be free from the neutrino oscillation data
as long as its contribution to the neutrino mass is negligible.
As found from eq.~(\ref{nmass}), such a situation can be realized for 
$|h_1|^2\Lambda_1 \ll |h_2|^2\Lambda_2$ in the present parameter setting.
This is good for the thermal leptogenesis since an 
appropriately small neutrino Yukawa coupling constant $\tilde h_{i1}^\nu$
makes both effective out-of-equilibrium decay of $N_{R_1}$ and
sufficient thermal production of the right-handed neutrino $N_{R_1}$ possible.

\subsection{Leptogenesis and DM abundance}
In this part, we proceed to the study of other phenomenological subjects 
such as leptogenesis and DM abundance.
Our main interest is what kind of results are obtained for these problems 
if we use the parameters assumed in the previous discussion. 
Since present model is defined even at larger scales than the
PQ symmetry breaking scale, we can also examine the consistency of the used
value of $\epsilon$ with the assumed symmetry breaking pattern in
(\ref{ssb}).

First, we discuss the leptogenesis in this model.
If we use the parameters assumed in the leptonic sector,
we can estimate baryon number asymmetry expected from
the out-of-equilibrium decay of the thermal $N_1$ by solving
the Boltzmann equation as done in \cite{ks}. 
The previous analysis in the similar model \cite{s,s1} shows that the required 
baryon number asymmetry could be generated for $M_1~{^>_\sim}~ 10^8$~GeV.
Since this value of $M_1$ is somewhat smaller than the Davidson-Ibarra
bound \cite{di} in the ordinary thermal leptogenesis \cite{cpv},
the reheating temperature could take a lower value than
usually assumed one to yield the thermal $N_{R_1}$.
This is crucial in the present model to forbid thermal production of
the extra colored fields $Q_{L,R}^{(i)}$ which cause dangerous relics
as discussed in the previous part.
If we assume the reheating temperature as $T_R \simeq M_1$,
we find $Y_B\sim 5\times 10^{-10}$ for the parameters given
in (\ref{rmass}) and (\ref{lyuk}), where $Y_B$ is defined as
$Y_B\equiv\frac{n_B}{s}$ by using the baryon number density $n_B$ 
and the entropy density $s$. In this calculation,
we assume a maximal $CP$ phase in the $CP$ violation parameter
$\varepsilon_1$ of the $N_{R_1}$ decay \cite{cpv} and  
an initial condition $Y_{N_1}(T_R)=0$ at the reheating 
temperature $T_R$.\footnote{We do not consider
  any additional $N_{R_1}$ production process other than the one caused by the
  neutrino Yukawa couplings. This is different from the analysis in \cite{kai}.
  As a result, we cannot make the mass of
  $N_1$ smaller than $10^8$~GeV for successful leptogenesis unless
  the degenerate right-handed neutrino masses are assumed.} 
Upper bound of the number density of the extra colored fermions
$Q_{L,R}^{(i)}$ might be estimated at $T_R$ by assuming that they are
in the thermal equilibrium. We find that the previously mentioned 
bound for $\frac{n_{Q^{(i)}}}{n_B}$ imposed by the search
for the fractionally charged particles could be satisfied for $Q^{(3)}$,
which has the smallest mass of $O(\epsilon^2 M_\ast)$ among 
the extra colored fermions.
Thus, the leptogenesis could be evaded from the dangerous relic problem
consistently.

Next, we address DM abundance in this model.
As mentioned before, the axion cannot be a dominant component 
of the DM in the scenario
since the upper bound of the decay constant required by the goodness
of the PQ symmetry is too small. However, the model has another DM candidate,
that is, the lightest neutral component of $\eta$ which is stable
because of $Z_2$ odd parity.
Its relic abundance is known to be controlled by the parameters
$\tilde\lambda_3$ and $\lambda_4$ in eq.~(\ref{scot}) since the
coannihilation among the components of $\eta$ is effective in case of
$\tilde m_\eta=O(1)$~TeV \cite{ks}.
In the left panel of Fig.~2, we plot typical points in the
$(\tilde\lambda_3, \lambda_4)$ plane, where the required DM 
abundance $\Omega_{\rm DM}h^2=0.12$ is realized by the relics of $\eta_R$.
In this calculation, we use $\tilde m_\eta=1$~TeV and 
$\tilde\lambda_5=5.4\times 10^{-3}$ 
which are assumed in the previous part.
In this panel, we take into account the condition $\lambda_4<0$ 
which is necessary for a neutral component of $\eta$ is lighter than
charged ones. If we use the tree level Higgs mass formula  
$m_{h^0}^2=4\tilde\lambda_1\langle\phi\rangle^2$, 
we find $\tilde\lambda_1\simeq 0.13$ for $m_{h^0}=125$~GeV.
This allows us to plot the last one in the stability condition (\ref{stab}) 
as a straight line in the same plane for a fixed $\tilde\lambda_2$. 
Points contained in the region above a straight line 
satisfy this condition for a fixed $\tilde\lambda_2$. 
Although the required DM abundance can be obtained for negative values of
$\tilde\lambda_3$, such cases contradict with the
condition for $\tilde\lambda_3$ given in eq.~(\ref{stab}).
The figure shows that $\tilde\lambda_3$ and/or $|\lambda_4|$ should 
take rather large values to realize the required DM abundance. 
Since they are used as initial values at the weak scale,
RG evolution of the scalar quartic couplings $\tilde\lambda_i$
could be largely affected.
In that case, vacuum stability and perturbativity of the model could 
give constraints on the assumed symmetry breaking scale $M_\ast$, which
should be smaller than a violation scale of vacuum stability
and perturbativity.
We focus our study on this point in the next part. 

\begin{figure}[t]
\begin{center}
\epsfxsize=7.5cm
\leavevmode
\epsfbox{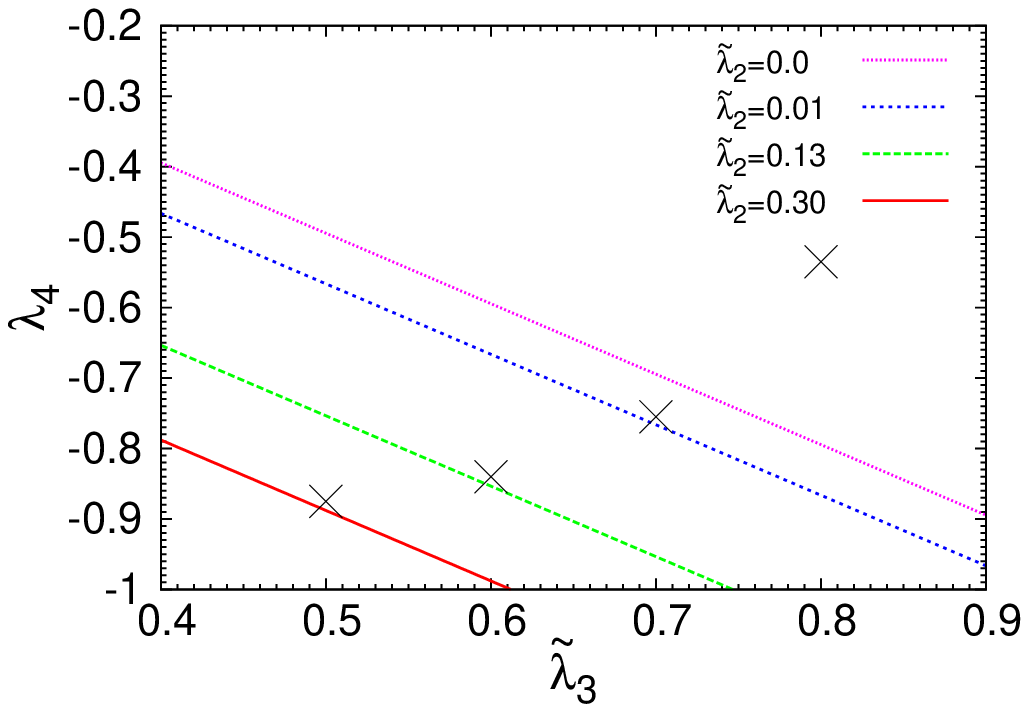}
\hspace*{5mm}
\epsfxsize=7.5cm
\leavevmode
\epsfbox{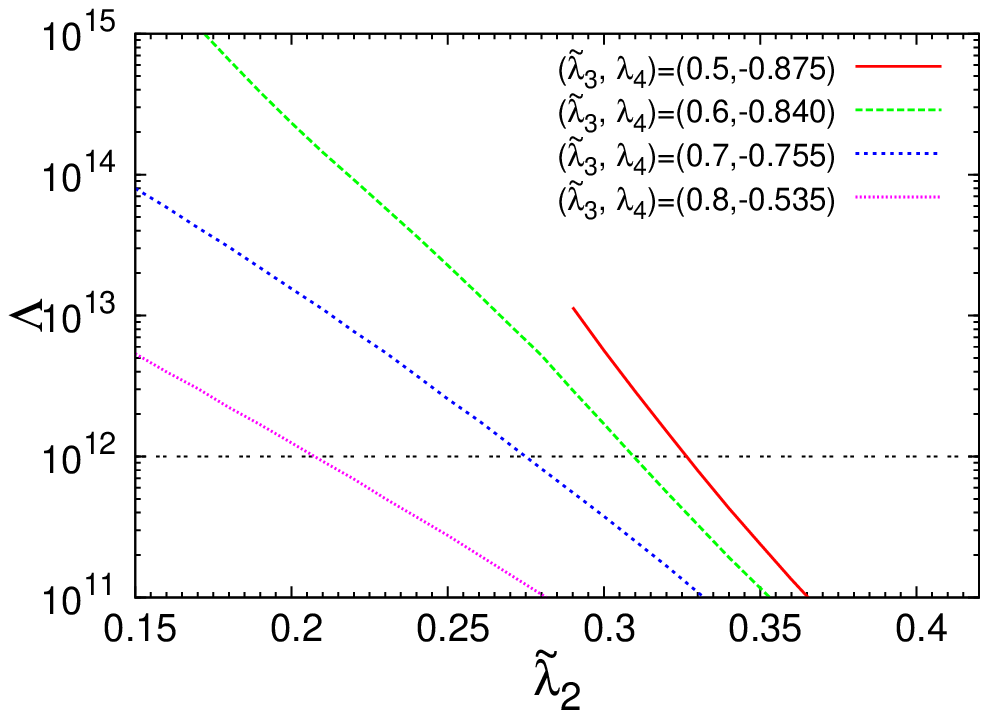}
\end{center}
\vspace*{-3mm}

{\footnotesize {\bf Fig.~2}~~
  Left panel: example points in the $(\tilde\lambda_3, \lambda_4)$ plane are
plotted by crosses, at which
  $\eta_R$ relics can explain the DM abundance $\Omega h^2=0.12$.
  $\tilde m_\eta^2=1$~TeV and $\tilde\lambda_5=5.4\times 10^{-3}$ are 
assumed. 
Above region of each line fixed by a listed value of $\tilde\lambda_2$
satisfies the last condition for the vacuum stability in (\ref{stab}).\\
Right panel: a cut-off scale $\Lambda$ as a function of $\tilde\lambda_2$,
which is fixed as a value at $M_Z$. Each line is plotted for four points marked 
by the crosses in the left panel where $\Omega h^2=0.12$ is satisfied.}
\end{figure}

\subsection{Symmetry breaking pattern and a cut-off scale}
We assume $\epsilon=0.07$ and $M_\ast=10^{12}$~GeV
in the previous part, which means
$\langle S\rangle= 7\times 10^{10}$~GeV.
It is crucial for the consistency of the scenario
whether $M_\ast$ is smaller than
a scale where either the vacuum stability or the
perturbativity is violated.\footnote{The constraint due to the vacuum stability 
and the perturbativity is taken into account in the DM study of 
the inert doublet model from a different viewpoint in \cite{inert1,inert2}.
The consistency between fermionic DM and the vacuum stability
is also studied in the scotogenic model \cite{stabf1,stabf2}.}  
We examine this problem by using the values of $\tilde\lambda_3$ and
$\lambda_4$ for which the required DM abundance is realized.  
Since the violation of the perturbativity is considered to suggest
a scale for the applicability of the model defined by eq.~(\ref{smodel}),
it should be larger than $M_\ast$.
This allows us to judge whether the $\epsilon$ value assumed
in the above phenomenological study is consistent with
the assumed symmetry breaking pattern.  

One-loop $\beta$-functions for the scalar quartic coupling constants in 
the effective model at energy regions below $M_S(\equiv\langle S\rangle)$
are given as follows \cite{rge}, 
\begin{eqnarray}
\beta_{\tilde\lambda_1}&=&24\tilde\lambda_1^2
+\tilde\lambda_3^2+(\tilde\lambda_3+\lambda_4)^2
+\tilde\lambda_5^2 \nonumber\\
&+&\frac{3}{8}\left(3g^4+g^{\prime 4}+2g^2g^{\prime 2}\right)
-3\tilde\lambda_1\left(3g^2+g^{\prime 2}-4h_t^2\right)-6h_t^4, \nonumber \\
\beta_{\tilde\lambda_2}&=&24\tilde\lambda_2^2+\tilde\lambda_3^2
+(\tilde\lambda_3+\lambda_4)^2+\tilde\lambda_5^2 \nonumber\\
&+&\frac{3}{8}\left(3g^4+g^{\prime 4}+2g^2g^{\prime 2}\right)
-3\tilde\lambda_2\left(3g^2+g^{\prime 2}\right), \nonumber \\
\beta_{\tilde\lambda_3}&=&2(\tilde\lambda_1+\tilde\lambda_2)
(6\tilde\lambda_3+2\lambda_4)
+4\tilde\lambda_3^2+2\lambda_4^2+2\tilde\lambda_5^2 \nonumber\\
&+&\frac{3}{4}\left(3g^4+g^{\prime 4}-2g^2g^{\prime 2}\right)
-3\tilde\lambda_3\left(3g^2+g^{\prime 2}-2h_t^2\right), \nonumber \\
\beta_{\lambda_4}&=&4(\tilde\lambda_1+\tilde\lambda_2)\lambda_4
+8\tilde\lambda_3\lambda_4+4\lambda_4^2
+8\tilde\lambda_5^2+3g^2g^{\prime 2}
-3\lambda_4\left(3g^2+g^{\prime 2}-2h_t^2\right), \nonumber\\
\beta_{\tilde\lambda_5}&=&4(\tilde\lambda_1+\tilde\lambda_2)\tilde\lambda_5
+8\tilde\lambda_3\tilde\lambda_5+12\lambda_4\tilde\lambda_5 
-3\tilde\lambda_5\left(3g^2+g^{\prime 2}-2h_t^2\right), 
\end{eqnarray}
where $\beta_\lambda$ is defined as 
$\beta_\lambda=16\pi^2\mu\frac{d\lambda}{d\mu}$ and the 
top Yukawa coupling is only taken into account among Yukawa interactions. 
In these equations, the positive contributions 
of $\tilde\lambda_3$ and $\lambda_4$ to the $\beta$-functions 
of $\tilde\lambda_{1,2}$ are found to tend to save the model
from violating the first condition in eq.~(\ref{stab}).
On the other hand, the same contributions of $\tilde\lambda_3$ and 
$\lambda_4$ could induce the violation of the perturbativity of the model 
at a rather low energy scale since they could give large positive 
contributions to $\beta_{\tilde\lambda_1}$, $\beta_{\tilde\lambda_2}$ and 
$\beta_{\tilde\lambda_3}$.
If we identify an applicable scale of the model defined
by eq.~(\ref{smodel}) with a scale $\Lambda$ 
where any of the perturbativity conditions $\lambda_i<4\pi$ and
$\kappa_i<4\pi$ is violated, $M_\ast <\Lambda$ should be satisfied.
If $M_\ast$ is larger than $\Lambda$, the consistency of 
the scenario is lost. 

We analyze this issue by solving the above one-loop RGEs at $\mu<M_S$
and also the ones which are given in \cite{s} at $\mu>M_S$.
The quartic couplings $\tilde\lambda_i$ 
in the tree-level potential at the energy scale $\mu<M_S$ are connected 
with the ones $\lambda_i$ at $\mu>M_S$ through eq.~(\ref{gcoupl1}).
Since the masses of the right-handed neutrinos $N_i$ are considered 
to be of $O(10^{8-9})$~GeV, they decouple 
at the scale $\mu<M_i~{^<_\sim}~O(M_S)$ to be irrelevant to the RGEs there.
On the other hand, the mass of the colored fields $Q_{L,R}^{(i)}$ are
required to be much heavier than $N_i$ as discussed before,
they can contribute mainly to the $\beta$-functions of the
$SU(3)_c$ gauge coupling at scales larger than their masses. 

The free parameters in the scalar potential of the low energy
effective model (\ref{scot}) 
are $\tilde\lambda_1,~\tilde\lambda_2,~\tilde\lambda_3,~\lambda_4$ and 
$\tilde\lambda_5$ at $M_Z$.\footnote{Quartic couplings $\kappa_i$ 
for $S$ are fixed as $\kappa_1=\frac{M_S^2}{4\langle S\rangle^2}$ 
and $\kappa_{2,3}=0.1$ at $M_S$ in this study.
Larger values of $\kappa_{2,3}$ make $\Lambda$ smaller.}
$\tilde\lambda_1$ is fixed at $\tilde\lambda_1\simeq 0.13$ from
the Higgs mass. Both $\tilde\lambda_3$ and $\lambda_4$ are
fixed at values determined through the DM relic abundance which are
shown in the left panel of Fig.~2. $\tilde\lambda_5$ is fixed at
$\tilde\lambda_5=5.4\times 10^{-3}$ which is used in the
discussion of the neutrino mass and the leptogenesis.
Thus, an only free parameter is $\tilde\lambda_2$.
If we solve the RGEs varying the value of $\tilde\lambda_2$,
we can search $\Lambda$ checking 
the vacuum stability and the perturbativity for each $\tilde\lambda_2$.

In the right panel of Fig.~2, we plot $\Lambda$ as a function 
of $\tilde\lambda_2$ for four sets of $(\tilde\lambda_3,~\lambda_4)$
which are shown by crosses in the left panel of Fig.~2.
An end point found in a line for $(0.5,-0.875)$ represents a value 
of $\tilde\lambda_2$
for which the vacuum stability is violated before reaching a scale of the 
perturbativity violation. 
This figure shows that $\Lambda$ could be high enough to be 
consistent with an assumed value of $\epsilon$ as long as 
$\tilde\lambda_2$ takes a suitable value. 
The present scenario for the symmetry
braking could be consistent with the explanation presented here for 
various phenomenological subjects.
The simultaneous explanation of the neutrino masses and the
DM abundance could be preserved in this extended model in the same way 
as in the original scotogenic model with heavy right-handed neutrinos.

\section{Summary}  
We have proposed a model which could solve the strong $CP$ problem
based on the PQ mechanism. The model is constructed to escape the domain wall
problem and to keep the goodness of the PQ symmetry against the breaking due
to the gravity effect. For this purpose, we introduce a local $U(1)_g$
symmetry and also a flavor dependent global $U(1)_{FN}$ symmetry.
The PQ symmetry is induced from these as their linear combination
through their spontaneous breaking.
Resulting PQ symmetry becomes flavor dependent to realize $N_{DW}=1$.
Its flavor dependence causes hierarchical masses and flavor mixing
for quarks and leptons after the PQ symmetry breaking.
Observed masses and flavor mixing seem to be
obtained in this framework without serious fine tuning for the coupling constants
of the nonrenormalizable operators. 
Moreover, after the $U(1)_{PQ}$ symmetry breaking, 
its subgroup $Z_2$ remains as a remnant exact symmetry at least 
in the leptonic sector.
So, the leptonic part of the model reduces to the well-known 
scotogenic model for the neutrino masses and DM, in which
the neutrino masses are generated through one-loop radiative effect 
and the DM abundance can be explained as the thermal relics of 
a neutral component of an extra doublet scalar.

The model can explain the cosmological baryon number asymmetry
through the out-of-equilibrium decay of a right-handed neutrino
in the same way as the ordinary thermal leptogenesis
in the tree-level seesaw model.
However, since the lower bound for the right-handed neutrino mass is 
relaxed in this model, the required reheating temperature could 
be low enough not to restore the PQ symmetry and also not to yield 
heavy colored particles in a dangerous amount in the thermal plasma. 
We also show that these features could be consistently
realized for suitable parameter sets.
Although we do not address inflation in this study, it might be
introduced into the model in the similar way discussed in \cite{infl}.
Since the simple extension discussed here can relate the strong $CP$
problem to the flavor structure of quarks and leptons, and
the origin of neutrino masses and DM, it may be promising
to consider an extended SM in this direction further.

\section*{Acknowledgements}
This work is partially supported by MEXT Grant-in-Aid 
for Scientific Research on Innovative Areas (Grant No. 26104009)
and a Grant-in-Aid for Scientific Research (C) from Japan Society
for Promotion of Science (Grant No. 18K03644).

\newpage
\bibliographystyle{unsrt}

\end{document}